\documentclass[conference]{IEEEtran}
\usepackage[a4paper, left=.67in, right=.71in, bottom=1in, top=0.8in]{geometry}
\IEEEoverridecommandlockouts
\usepackage[nospace,noadjust]{cite}
\usepackage[dvipsnames]{xcolor}
\usepackage{amsmath,amssymb,amsfonts}
\usepackage{amsthm}
\usepackage{graphicx,caption}
\usepackage{bbm}
\usepackage{setspace}
\usepackage{braket}
\usepackage{algpseudocode}
\usepackage[]{algorithm} 
\usepackage{textcomp}
\usepackage{xcolor} 
\usepackage{soul}
\usepackage{caption}
\def\BibTeX{{\rm B\kern-.05em{\sc i\kern-.025em b}\kern-.08em
		T\kern-.1667em\lower.7ex\hbox{E}\kern-.125emX}}
\usepackage{nccmath}
\usepackage{array}
\usepackage{subfig}
\usepackage{bigstrut}
 \usepackage{booktabs}
\definecolor{wblue}{RGB}{0,58,199}
\usepackage{setspace}
\usepackage{multicol}
\usepackage[T1]{fontenc}
\usepackage[utf8]{inputenc}
\usepackage{babel}
\usepackage[font=small,labelfont=bf]{caption}
\usepackage{bm}
\usepackage{blindtext}
\usepackage{caption}
\usepackage{bigstrut}
\algnewcommand{\IIf}[1]{\State\algorithmicif\ #1\ \algorithmicthen}
\algnewcommand{\EndIIf}{\unskip\ \algorithmicend\ \algorithmicif}
\algnewcommand{\FFor}[1]{\State\algorithmicfor\ #1\ \algorithmicdo}
\algnewcommand{\EndFFor}{\unskip\ \algorithmicend\ \algorithmicfor}

\ifCLASSINFOpdf
\else
\fi

\begin{document}
%
\title{
	{Channel Reuse for Backhaul in UAV Mobile Networks with User QoS Guarantee}
	{\footnotesize \textsuperscript{}}}

	\author{\IEEEauthorblockN{Mohammadsaleh Nikooroo\textsuperscript{1}, Zdenek Becvar\textsuperscript{1}, Omid Esrafilian\textsuperscript{2},  David Gesbert\textsuperscript{2}}
	\IEEEauthorblockA{\textit{\textsuperscript{1} Faculty of Electrical Engineering} 
		\textit{Czech Technical University in Prague},
		Prague, Czech Republic \\
		\textsuperscript{2}	\textit{Communication Systems Department, EURECOM}, Sophia Antipolis, France\\
		\textsuperscript{1}\{nikoomoh,zdenek.becvar\}@fel.cvut.cz, \textsuperscript{2}\{esrafili,gesbert\}@eurecom.fr }
	\thanks{\textcolor{black}{This work was supported by the project No. LTT 20004 funded by Ministry of Education, Youth and Sports, Czech Republic, and by the grant of Czech Technical University in Prague No. SGS20/169/OHK3/3T/13, and partially by the HUAWEI
France supported Chair on Future Wireless Networks at EURECOM.}}}

\maketitle

\begin{abstract}
	In mobile networks, unmanned aerial vehicles (UAVs) acting as flying base stations (FlyBSs) can effectively improve performance. Nevertheless, such potential improvement requires an efficient positioning of the FlyBS. In this paper, we study the problem of sum downlink capacity maximization in FlyBS-assisted networks with mobile users and with a consideration of wireless backhaul with channel reuse while a minimum required capacity to every user is guaranteed. The problem is formulated under constraints on the FlyBS's flying speed, propulsion power consumption, and transmission power for both of flying and ground base stations. None of the existing solutions  maximizing the sum capacity can be applied due to the combination of these practical constraints. This paper pioneers in an inclusion of all these constraints together with backhaul to derive the optimal 3D positions of the FlyBS and to optimize the transmission power allocation for the channels at both backhaul and access links as the users move over time. The proposed solution is geometrical based, and it shows via simulations a significant increase in the sum capacity (up by 19\%-47\%) compared with baseline schemes where one or more of the aspects of backhaul communication, transmission power allocation, and FlyBS's positioning are not taken into account.  
\end{abstract}


\section{Introduction}\label{sec:1}
\par

Unmanned aerial vehicles (UAVs) have received an extensive attention in wireless communications in the recent years. Due to a high flexibility and adaptability to the environment, UAVs can be regarded as flying base stations (FlyBSs) that potentially bring a significant enhancement in the performance of mobile networks \cite{Esrafilian}. Such potential enhancements, however, are essentially subject to an effective management of several aspects, including propulsion power consumption, transmission power consumption/allocation,  FlyBS's positioning , etc. As another crucial aspect, a backhaul communication of the FlyBSs with the ground base station (GBS) or access point (AP) must be ensured  in order to integrate FlyBSs into mobile networks.


 Many recent work investigate the performance in FlyBS-assisted networks with inclusion of backhaul.  In \cite{Cicek2020}, they address the FlyBS’s positioning and bandwidth allocation to optimize the total profit gained from the users in a network. Furthermore, the authors in \cite{Pan2019} investigate an optimization of the FlyBS’s position, user association, and resource allocation, to maximize the utility in software-defined cellular networks.  In \cite{Li2019}, they maximize energy efficiency  in a relaying network with static BSs via optimization of transmission power allocation to the BSs. Then, the problem of joint 2D trajectory design and resource allocation is investigated in \cite{Yu2021} to minimize the network latency in a space-air-ground network with millimeter wave (mmWave) backhaul.
In \cite{Qiu2020TCOM}, the authors study a joint placement, resource allocation, and user association of FlyBSs to maximize the network’s utility. Then, in \cite{Huang2020}, they  maximize the minimum rate of the delay-tolerant users via a joint resource allocation and FlyBS’s positioning. Then, in \cite{Liu2021}, the authors consider a scenario where a set of relaying FlyBSs establish a communication between multiple sources and multiple destinations. The goal is to maximize the minimum average rate among the relays via transmission power allocation and FlyBSs' positioning. Furthermore, in \cite{LZhang2021}, the operation cost in a mobile edge computing network is minimized via FlyBS's positioning and resource allocation. 
The solutions provided in \cite{Cicek2020}-\cite{LZhang2021} do not assume constraints on the user's instantaneous capacity and, hence, they cannot be applied in scenarios with delay-sensitive users where a minimum  capacity is demanded by the users. 


Several works also consider the individual user’s quality of service in terms of instantaneous capacity. In \cite{Youssef2020} and \cite{Kalantari2020}, the problem of FlyBS positioning and resource allocation is investigated to minimize the transmission power of the FlyBS. Then, the minimum capacity of the users is maximized via the FlyBS’s positioning and the transmission power allocation in \cite{Valiulahi2020}. Furthermore, the problem of transmission power allocation is investigated  in \cite{Muntaha2021} for FlyBS networks to maximize the energy efficiency, i.e., the ratio of the sum  capacity to the total transmission power consumption. 
The authors in \cite{Sabzehali2022} minimize the number of FlyBSs in a network while ensuring both coverage to all ground users. Then, in \cite{Qiu2020Access}, the problem of resource allocation and circular-trajectory design for fixed-wing FlyBSs is investigated to minimize the power consumption of the FlyBS.  Furthermore, the minimum capacity of the users is maximized in \cite{Iradukunda2021} via resource allocation and positioning. Also,  the minimum downlink throughput is maximized in \cite{Li2018} by optimizing the FlyBSs’ positioning, bandwidth, and power allocation.


In our prior work \cite{Nikooroo_Globecom_2022}, the problem of FlyBS's positioning and user association is investigated in mobile networks assisted by relaying FlyBSs. Also, in \cite{Nikooroo_Pimrc_2022} and \cite{Nikooroo2022TNSE}, a positioning of the FlyBS and transmission power allocation is proposed at the access link to maximize the sum capacity and to minimize the FlyBS's total power consumption, respectively, where a minimum required capacity for the users is guaranteed.

To our best knowledge, there is no work targeting the sum capacity maximization in a practical scenario with \textit{moving users} and with the \textit{minimum capacity guaranteed to the individual users} where a \textit{backhaul link}  is also provided. 
All related works either target scenario where no minimum capacity is guaranteed to the users and/or a backhaul connection (together with related backhaul constraints \cite{Youssef2020}, \cite{Mach2022}) is missing. It is also noted that, existing solutions maximizing the minimum capacity among users cannot be applied in many scenarios where users require different instantaneous capacities. To this end, we target the case with both backhaul and user's required capacity and we propose an analytical solution based on an alternating optimization of the FlyBS’s positioning and the transmission power allocation at the backhaul and at the access links. Due to a non-convex nature of the problem, a heuristic solution is proposed with respect to the feasibility region that is determined via constraints in the problem, i.e., \textcolor{black}{1) user’s required capacity at all time, 2) FlyBS's maximum speed, 3)  maximum propulsion power consumption of FlyBS, and 4) flow conservation constraint regarding the backhaul and access links.}



\section{System model and problem formulation}\label{sec:2}

In this section, we first define the system model. Then, we formulate the constrained problem of sum capacity maximization with inclusion of backhaul communication.

\begin{figure}[!t]
	\centering
	\includegraphics[width=3.5in]{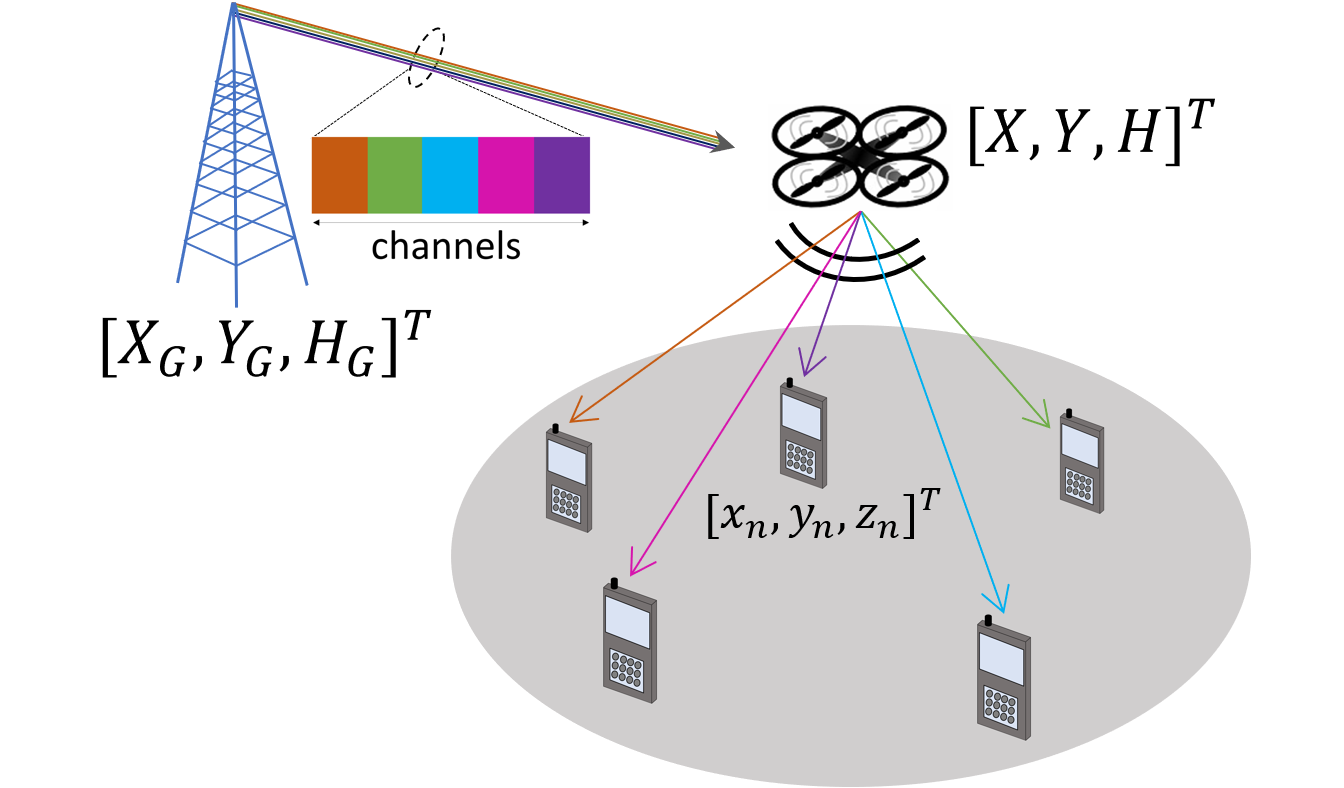}	
	\captionsetup{justification=centering}
	\caption{System model with mobile users placed within the coverage area of the FlyBS. The channels at the access link are reused for GBS-to-FlyBS communication} 
	\label{fig:sysmodel}
\end{figure}

In our system model, the FlyBS serves $ N $ mobile users in an area as shown in Fig. \ref{fig:sysmodel}. The FlyBS connects to the GBS located at $ \bm{l_G}=[X_G, Y_G, H_G] $ via backhaul. Let , $ \bm{l_F}[k]=\big[X[k], Y[k], H[k]\big]^T $  and $ \bm{u_n}[k]=\big[x_n[k],\ y_n[k],z_n[k]\big]^T $ denote the location of the FlyBS and the user $n$ at the time step $ k $, respectively. Also, let $ d_{n,F}[k] $ and $ d_{n,G}[k] $ denote the Euclidean distance of the user  $ u_n $ to the FlyBS and to the GBS at the time step $ k $, respectively.

	 Suppose the whole available radio band is divided into a set of $S$ channels $ \textbf{J}=\{J_1,…,J_S\} $, where channel $ J_s $ has a bandwidth of $ B_s $ $ (1\leq s\leq S) $. At the FlyBSs, we adopt orthogonal downlink channel allocation to all users. Furthermore, all the $S$ channels are reused at the backhaul link to alleviate the scarcity of  radio resources. Let $ g_n \in [1,S]$  be the index of the channel allocated to user $ n $. Note that, we do not target an optimization of channel allocation in this paper, and we leave that for future work. Nevertheless, our model works with any channel allocation.

 Let $p_{n,F}^R$ be the received power at the user $n$ from the FlyBS. Furthermore, $p_{F,G,s}^R$ denotes the received power at the FlyBS from the GBS over the channel $s$. Then, the channel capacity of the user $ n $ is:
\begin{gather}
C_n[k]=B_{g_n}\log_2{\left(1+\frac{p_{n,F}^R[k]}{\sigma^2_n+p_{n,G}^R[k]}\right)},
\label{eqn:1}
\end{gather}

\noindent where $p_{n,G}^R$ is the interference power received at user $n$ from the GBS, $\sigma^2_n$ is noise's power. Similarly, the link's capacity between the GBS and the FlyBS is:
\begin{gather}
C_{G,F}[k]=\sum_{s=1}^S B_{s}\log_2{\left(1+\frac{p_{F,G,s}^R[k]}{\sigma^2_{F,s}}\right)},
\label{eqn:2}
\end{gather}
where $\sigma^2_{F,s}$ is the noise power over the channel $s$. 

Let  $ \bm{p_F}^T=[p_{F,1}^T,...,p_{F,N}^T] $ denote the FlyBS’s transmission power vector to all the users. Also, for the GBS-to-FlyBS communication, let $ \bm{p_G}^T=[p_{G,1}^T,...,p_{G,S}^T] $ be the GBS's transmission power vector over the $S$ channels. According to the Friis’ transmission equation, we have 
\begin{gather}\label{eqn:friis_tx_rx_Fn}
p_{n,F}^R=Q_{n,F}p_{F,n}^T{d_{n,F}}^{-\alpha_{n,F}}, n\in [1,N],
\end{gather}
\noindent where the coefficient $ Q_{n,F} $ is the parameter depending on the communication frequency and the gain of antennas, and $ \alpha_{n,F} $ is the pathloss exponent of the channel between the FlyBS and the user $n$. Similar relation can be derived between the GBS's transmission power and the received power at the user $n$ and at the FlyBS as
\begin{gather}
p_{n,G}^R=Q_{n,G}p_{G,g_n}^T{d_{n,G}}^{-\alpha_{n,G}},n\in [1,N],\label{egn:friis_tx_rx_G}\\
p_{F,G}^R=Q_{F,G}p_{G,s}^T{d_{F,G}}^{-\alpha_{F,G}}, n\in [1,N], s\in [1,S].\nonumber
\end{gather}


\textcolor{black}{For the propulsion power consumption, we refer to the model provided in \cite{Zeng2019} for rotary-wing UAVs, where the propulsion power is expressed as:}
\textcolor{black}{{\begin{gather}
P_{pr}[k]=L_{0}\big(1+\frac{3V_{F}^{2}[k]}{U_{\text{tip}}^{2}}\big)+\frac{\eta_0\rho s_rAV_{F}^3[k]}{2}+\nonumber\\L_{i}\big(\sqrt{1+\frac{V_{F}^4[k]}{4v^4_{0,h}}}-\frac{V_{F}^{2}[k]}{2v^2_{0,h}}\big)^{\frac{1}{2}},\label{eqn:PropulsionModel}
\end{gather}}
}

\noindent \textcolor{black}{where $V_F[k]$ is the FlyBS's speed at the time step $k$, $ L_0 $ and $ L_i $ are the blade profile and induced powers in hovering status, respectively, $ U_{\text{tip}} $ is the tip speed of the rotor blade, $ v_{0,h} $ is the mean rotor induced velocity during hovering, $ {\eta_0} $ is the fuselage drag ratio, $ \rho $ is the air density, $ s_r $ is the rotor solidity,  and $ A $ is the rotor disc area. }

 \vspace{0\baselineskip}

Our goal is to find the optimized position of the FlyBS and to determine the transmission power allocation over each channel both at the backhaul and at the access link to maximize the sum capacity at every time step $ k $ under practical constraints as follows:

\begin{gather}
\operatorname*{max}_{\bm{p_G}^T[k],\bm{p_F}^T[k], \bm{l_F}[k]}{\;} \sum_{n=1}^{N}C_n[k],\quad\quad\quad\quad\quad \forall k,\label{eqn:problem_formulation_original}\\[+1pt]
\text{s.t.}\quad C_n[k]\geq C_{n, min}[k],  n\in [1,N],  \quad  (\text{\ref{eqn:problem_formulation_original}a})\nonumber\\[+3pt]
H_{min}[k]\leq H[k]\leq H_{max}[k], \quad\quad\quad\quad (\text{\ref{eqn:problem_formulation_original}b})\nonumber\\[+3pt]
\big|\big|\bm{l}[k]-\bm{l}[{k-1}]\big|\big|\le V_{F,max}\delta_k,\quad\quad\quad{\;}  (\text{\ref{eqn:problem_formulation_original}c})\nonumber\\[+3pt]
P_{pr}[k]\leq P_{pr,th}[k],\quad\quad\quad\quad\quad\quad\quad\quad{\;} (\text{\ref{eqn:problem_formulation_original}d})\nonumber\\[+1pt]
\sum\nolimits_{n=1}^{N}C_n[k]\leq C_{G,F}[k], \quad\quad\quad\quad\quad{\;}{\;} (\text{\ref{eqn:problem_formulation_original}e})\nonumber\\[+1pt]
\sum\nolimits_{n=1}^{N}p_{G,n}^T\le p_{G,max}^T,\quad  p_{G,n}^T\geq0 \quad{\;}{\;} (\text{\ref{eqn:problem_formulation_original}f}) \nonumber\\[+1pt]
\sum\nolimits_{s=1}^{S}p_{F,s}^T\le p_{F,max}^T,\quad p_{F,s}^T\geq 0  \quad{\;}{\;}{\;}{\;} (\text{\ref{eqn:problem_formulation_original}g}) \nonumber
\end{gather}

\noindent where $\delta_k$ is the duration between the time steps $k-1$ and $k$,  and $ ||.|| $ is the $  \mathcal{ L}_2 $ norm. The constraint (\ref{eqn:problem_formulation_original}a) ensures that every user always receives their minimum required capacity $ C_{n,min}[k] $. The constraint (\ref{eqn:problem_formulation_original}b) restricts the FlyBS's altitude within $[H_{min},H_{max}]$ where $H_{min}$ and $H_{max}$ are the minimum and maximum allowed flying altitude, respectively, and are set according to the environment and also the flying regulations. The constraint (\ref{eqn:problem_formulation_original}c) ensures the FlyBS's speed would not exceed the maximum supported speed $ V_{F,max} $, and (\ref{eqn:problem_formulation_original}d) assures that the FlyBS's movement would not incur the propulsion power larger than a threshold $P_{pr,th}$. 
In practice, the value of $P_{pr,th}$ can be set arbitrarily at every time step and according to available remaining energy in the FlyBS's battery to prolong the FlyBS's operation. 
Furthermore, (\ref{eqn:problem_formulation_original}f) and (\ref{eqn:problem_formulation_original}g) limit the total transmission powers of the GBS and the FlyBS to the maximum values of $p_{G,max}^T$ and $p_{F,max}^T$, respectively.  

In the next section, we elaborate on our proposed solution to the formulated problem in (\ref{eqn:problem_formulation_original}).



\section{FlyBS positioning and transmission power allocation on access and backhaul links}\label{sec:3}

In this section, we present our proposed solution to (\ref{eqn:problem_formulation_original}). We provide a high level overview of the optimization of the transmission power allocation on both access and backhaul links as well as the FlyBS's positioning. Then, we describe in details individual steps of the optimization in following subsections.

\subsection{Overview of the proposed solution}

Solving (\ref{eqn:problem_formulation_original}) in general is challenging, since the objective (i.e., sum capacity) is only convex with respect to $\bm{p_G}^T[k]$, as it is concave with respect to $\bm{p_F}^T[k]$ and also not convex (nor concave) with respect to $\bm{l_F}$. In addition, the constraints (\ref{eqn:problem_formulation_original}a), (\ref{eqn:problem_formulation_original}d), and (\ref{eqn:problem_formulation_original}e) are also not convex (nor concave) with respect to $\bm{l_F}$. 
Therefore, we propose a solution based on an alternating optimization of the power allocation and the FlyBS's positioning. In particular, the optimization in (\ref{eqn:problem_formulation_original}) is done via iterating the following three steps: 1) optimize $p_F^T$ at a given position of the FlyBS $q_F$ and for a fixed power allocation $p_G^T$, 2) optimize $p_G^T$ at the same given position of the FlyBS in \textit{step 1} and for the updated $p_F^T$ from \textit{step 1}, 3) optimize the FlyBS's position $ \bm{l_F}$ for the updated power allocation derived from \textit{steps 1} and \textit{2}. Furthermore, to tackle the non-convexity of the objective, we propose an approximation form of the objective that intuits us to what direction for the FlyBS's movement incurs an increase in the sum capacity. The idea of step-wise solving of  (\ref{eqn:problem_formulation_original}) facilitates to deal with the non-convexity of the constraints. Each step is solved with respect to the related constraints in  (\ref{eqn:problem_formulation_original}). In the next section, we elaborate our proposed solution. 

\subsection{Transmission power allocation for access link}

At a fixed position of the FlyBS and for a given setting of transmission power at the backhaul link ($\bm{p_G}^T$),  the problem of transmission power optimization at the access link to maximize the sum capacity is formulated as follows:

\begin{gather}
\operatorname*{max}_{\bm{p_F}^T[k]}{\;} \sum_{n=1}^{N}C_n[k],\quad\quad\quad\quad\quad \forall k,\label{eqn:opt_step2}\\[-0pt]
\text{s.t.}\quad (\text{\ref{eqn:problem_formulation_original}a}), (\text{\ref{eqn:problem_formulation_original}e}), (\text{\ref{eqn:problem_formulation_original}g}).\nonumber
\end{gather}
\noindent Note that, only constraints from (\ref{eqn:problem_formulation_original}) that directly relate to the optimization variable $\bm{p_F}^T$ are included in (\ref{eqn:opt_step2}). According to (\ref{eqn:1}) and (\ref{eqn:friis_tx_rx_Fn}), the objective in (\ref{eqn:opt_step2}) is concave and the constraint (\ref{eqn:problem_formulation_original}g) is convex with respect to $\bm{p_F}^T[k]$. Furthermore, the constraint (\ref{eqn:problem_formulation_original}a) is rewritten as 
\begin{gather}\label{eqn:6a_accesslink}
B_{g_n}\log_2{\left(1+\frac{Q_{n,F}p_{F,n}^T}{{d_{n,F}}^{\alpha_{n,F}}(\sigma^2_n+ p_{n,G}^R[k])}\right)}\geq C_{n,min}[k],
\end{gather}

\noindent or equivalently

\begin{gather}
p_{F,n}^T\geq Q^{-1}_{n,F}({2^{\frac{C_{n,min}[k]}{B_{g_n}}}-1})(\sigma^2_n+ p_{n,G}^R[k]){d_{n,F}}^{\alpha_{n,F}},\label{eqn:(5a)_equivalent_for_access}
\end{gather}

\noindent which is linear with respect to $p_{F,n}^T$. Next, we rewrite (\ref{eqn:problem_formulation_original}e) by the means of (\ref{eqn:1}) and (\ref{eqn:friis_tx_rx_Fn}) as:  
\begin{gather}
\sum_{n=1}^{N}B_{g_n}\log_2{\left(1+\frac{Q_{n,F}p_{F,n}^T}{d_{n,F}^{\alpha_{n,F}}(\sigma^2_n+p_{n,G}^R[k])}\right)}\leq C_{G,F}[k],\label{eqn:5e_for_access}
\end{gather}
which is non-convex with respect to $p_{F,n}^T$. To tackle this issue, we consider the inequality $\text{log}_2(1+a\mathcal{X})\leq \text{log}_2(a)+\frac{1}{\text{ln}(2)}(\text{ln}(\frac{1+s\tau}{a})+\frac{a\mathcal{X}-s\tau}{1+s\tau})$ for arbitrary values $a$ and $\mathcal{X}$, where $s=\lfloor\frac{a\mathcal{X}}{\tau}\rfloor$, and $\tau$ is an approximation
parameter in the Taylor series and choosing a smaller $\tau$ leads to a smaller gap between the two sides of the mentioned inequality. By adopting $s_n=\lfloor\frac{p_{F,n}^T}{\tau}\rfloor$ and by applying the inequality to the left-hand side in (\ref{eqn:5e_for_access}), we get:
\begin{gather}
\sum_{n=1}^{N}B_{g_n}\log_2{\left(1+\frac{Q_{n,F}p_{F,n}^T}{d_{n,F}^{\alpha_{n,F}}(\sigma^2_n+p_{n,G}^R[k])}\right)}\leq \label{eqn:5e_LHS_bound}\\
C_{\text{tot}}^{\text{ub}}=\sum_{n=1}^{N}B_{g_n}({Q^{-1}_{n,F}d_{n,F}^{\alpha_{n,F}}(\sigma^2_n+p_{n,G}^R[k])})+\nonumber\\
 \sum_{n=1}^{N}\frac{B_{g_n}}{\text{ln}(2)}(\text{ln}({Q^{-1}_{n,F}d_{n,F}^{\alpha_{n,F}}(\sigma^2_n+p_{n,G}^R[k])}(1+s_n\tau))-\nonumber
 \end{gather}
 \begin{gather}
 \frac{s_n\tau_n}{1+s_n\tau}+
\frac{p_{F,n}^T}{(1+s_n\tau)(\sigma^2_n+p_{n,G}^R[k])}).\nonumber
\end{gather}

\noindent $C_{\text{tot}}^{\text{ub}}$ substitutes the upper bound for sum capacity in the right-hand side in (\ref{eqn:5e_LHS_bound}).  Using (\ref{eqn:5e_LHS_bound}), we replace (\ref{eqn:problem_formulation_original}e) by the linear (with respect to $p_{F,n}^T$) inequality $ C_{\text{tot}}^{\text{up}}\leq C_{G,F}[k]$. Then, the problem in (\ref{eqn:opt_step2}) is solved using CVX.

\subsection{Transmission power allocation for backhaul link}
Once the power allocation $\bm{p_G}^T[k]$ over the access link channels is optimized, we optimize the power allocation over the backhaul. To this end, we derive the subproblem of the transmission power optimization at the access link from (\ref{eqn:problem_formulation_original}) as  
\begin{gather}
\operatorname*{max}_{\bm{p_G}^T[k]}{\;} \sum_{n=1}^{N}C_n[k],\quad\quad\quad\quad\quad \forall k,\label{eqn:opt_step1}\\[-0pt]
\text{s.t.}\quad (\text{\ref{eqn:problem_formulation_original}a}), (\text{\ref{eqn:problem_formulation_original}e}),\nonumber (\text{\ref{eqn:problem_formulation_original}f}).
\end{gather}

\noindent From (\ref{eqn:1}) and (\ref{egn:friis_tx_rx_G}), we observe that the objective as well as (\ref{eqn:problem_formulation_original}f) are convex with respect to $\bm{p_G}^T$. Furthermore, the constraint (\ref{eqn:problem_formulation_original}a) is rewritten similarly as for (\ref{eqn:6a_accesslink}) and (\ref{eqn:(5a)_equivalent_for_access}) as


\begin{gather}
p_{G,g_n}^T\leq (\frac{p_{n,F}^R[k]}{2^{\frac{C_{n,min}[k]}{B_{g_n}}}-1}-\sigma^2_n)Q^{-1}_{n,G}{d_{n,G}}^{\alpha_{n,G}}, \label{eqn:(5a)_equivalent_for_backhaul}
\end{gather}
which is linear with respect to $p_{G,g_n}^T$. Next, we rewrite (\ref{eqn:problem_formulation_original}e) by the means of (\ref{eqn:1}), (\ref{eqn:friis_tx_rx_Fn}), and (\ref{egn:friis_tx_rx_G}) as:  
\begin{gather}
\sum\nolimits_{n=1}^{N}B_{g_n}\log_2{\left(1+\frac{Q_{n,F}p_{F,n}^T}{d_{n,F}^{\alpha_{n,F}}(\sigma^2_n+\frac{Q_{n,G}p_{G,g_n}^T}{d_{n,G}^{\alpha_{n,G}}})}\right)}-\nonumber
\end{gather}
\begin{gather}
\sum_{s=1}^S B_{s}\log_2{\left(1+\frac{Q_{F,G}p_{G,s}^T[k]}{\sigma^2_{F,s}d_{F,G}^{\alpha_{F,G}}}\right)}\leq 0,
\end{gather}
which is convex with respect to $p_{G,g_n}^T$ for $p_{G,g_n}^T\ge 0$. Hence, the problem in (\ref{eqn:opt_step1}) is a concave programming problem. Similar to the convex optimization, efficient solutions are developed  in the literature for such class of problems in case that constraints define a convex compact set (like in (\ref{eqn:opt_step1})). 

We develop a solution based on an iterative construction of level sets for the objective function and derivation of local solutions with respect to the level sets using linear programming (LP), see \cite{Chinchuluun2005}.

\subsection{FlyBS positioning}

After optimizing the power allocation over the access and backhaul channels, we propose a solution to the FlyBS's positioning. To this end, we formulate the problem as: 
\begin{gather}
\operatorname*{max}_{\bm{l_F}[k]}{\;} \sum_{n=1}^{N}C_n[k],\quad\quad\quad\quad\quad \forall k,\label{eqn:opt_step3}\\[-0pt]
\text{s.t.}\quad (\text{\ref{eqn:problem_formulation_original}a}), (\text{\ref{eqn:problem_formulation_original}b}), (\text{\ref{eqn:problem_formulation_original}c}), (\text{\ref{eqn:problem_formulation_original}d}), (\text{\ref{eqn:problem_formulation_original}e}).\nonumber
\end{gather}
The objective and the constraints (\ref{eqn:problem_formulation_original}c) and (\ref{eqn:problem_formulation_original}e) are not convex with respect to $\bm{l_F}[k]$. Before dealing with the mentioned non-convexity, let us first discuss the constraints (\ref{eqn:problem_formulation_original}a), (\ref{eqn:problem_formulation_original}c), and (\ref{eqn:problem_formulation_original}d).\newline


\noindent\textit{C.1) Interpretation of constraints}\vspace{+2mm}

  The constraint (\ref{eqn:problem_formulation_original}a) is rewritten as  
\begin{gather}\label{eqn: 5a_in_positioning_subproblem}
{d_{n,F}}\leq (\frac{Q_{n,F}p_{F,n}^T}{(2^{\frac{C_{n,min}[k]}{B_{g_n}}}-1)(\sigma^2_n+ p_{n,G}^R[k])})^{\frac{1}{\alpha_{n,F}}}, \quad \forall n
\end{gather}
which defines as the FlyBS's next possible position as the border and inside of a sphere with a center at $\bm{u_n}[k]$ and with a radius of the right-hand side in \ref{eqn: 5a_in_positioning_subproblem}.

\begin{figure}[!t]
	\centering
	\includegraphics[width=2.4in]{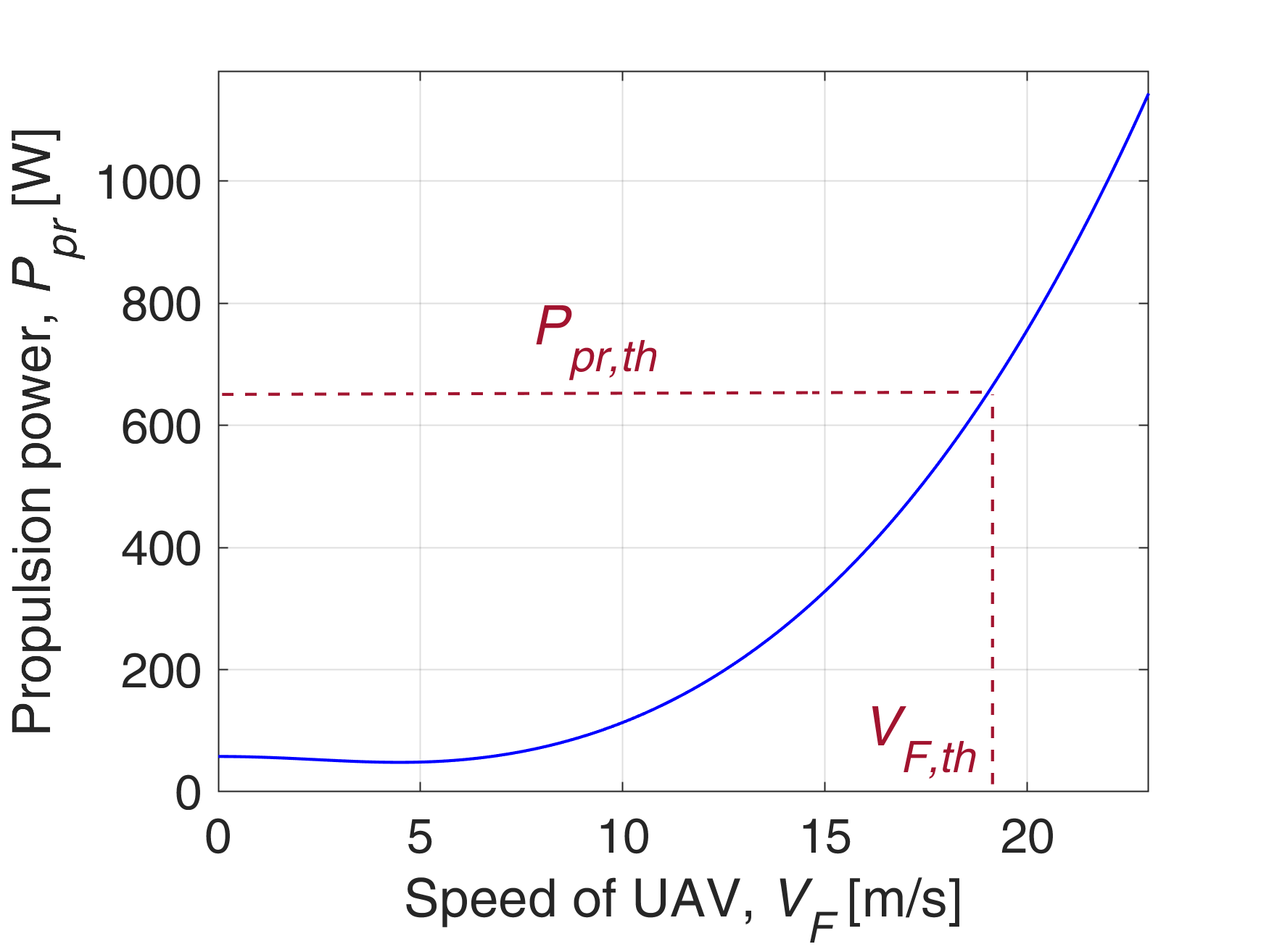}	
	\caption{\textcolor{black}{Propulsion power model vs. speed for rotary-wing FlyBS. }} 
	\label{fig:propulsion_model}
\end{figure}

According to Fig. \ref{fig:propulsion_model} the constraint (\ref{eqn:problem_formulation_original}d) is equivalent to $V_F$ being upper bounded by a threshold $ V_{F,th}$, i.e.,  $V_F\leq V_{F,th}$. By combining this inequality with (\ref{eqn:problem_formulation_original}c) we get 
	\begin{gather}
		||\bm{l}[k]-\bm{l}[k-1]||\le (\text{min}\{V_{F,max},V_{F,th}\})\delta_k, \label{eqn:4c-4d_combined_1}
	\end{gather}

\noindent Equation (\ref{eqn:4c-4d_combined_1}) defines the FlyBS’s next possible position as the border or inside of the region enclosed by two spheres centered at $ \bm{l}[k-1] $ (i.e., the FlyBS’s position at the previous time step), one with a radius of  $ V_{F,th}\delta_k$ and the other one with ($\text{min}\{V_{F,max},V_{F,th}\})\delta_k$.

Next, to deal with the non-convexity in (\ref{eqn:problem_formulation_original}e), let us first derive an upper bound for the left-hand side in (\ref{eqn:problem_formulation_original}e). To this end, we use the fact the FlyBS's next position is bounded due to the limit on the FlyBS's speed as well as altitude. More specifically, from (\ref{eqn:4c-4d_combined_1}) we find a lower bound to the FlyBS's distance from user $n$ ($n \in [1,N]$) at time step $k$ in terms of the FlyBS's position at time step $k-1$ as
\begin{gather}
		d_{n,F}[k]\geq d_{n,F,min}[k]= \text{max}\{H_{min}[k],\label{eqn:distance_lower_bound}\\	||\bm{l_F}[k-1]-\bm{u_n}[k]||+(\text{min}\{V_{F,max},,V_{F,th}\})\delta_k\},\nonumber
\end{gather}

\noindent Then, by using (\ref{eqn:distance_lower_bound}), we get the following upper bound for the left-hand side in (\ref{eqn:problem_formulation_original}e):
\begin{gather}
\sum_{n=1}^{N}C_n[k]\leq \label{eqn:sum_C_lower_bound}\\ \sum_{n=1}^{N}B_{g_n}\log_2{\left(1+\frac{Q_{n,F}p_{F,n}^T}{{d_{n,F,min}}^{\alpha_{n,F}}(\sigma^2_n+ p_{n,G}^R[k])}\right)},\nonumber
\end{gather}

\noindent Thus, we replace (\ref{eqn:problem_formulation_original}e) with the following constraint:

\begin{gather}
 \sum_{n=1}^{N}B_{g_n}\log_2{\left(1+\frac{Q_{n,F}p_{F,n}^T}{{d_{n,F,min}}^{\alpha_{n,F}}(\sigma^2_n+ p_{n,G}^R[k])}\right)}\leq\nonumber
 \end{gather}
 \begin{gather}
C_{G,F}[k]=\sum_{s=1}^S B_{s}\log_2{\left(1+\frac{Q_{F,G}p_{G,s}^T[k]}{\sigma^2_{F,s}d_{F,G}^{\alpha_{F,G}}}\right)}. \label{eqn:5e_equivalent_positioning}
\end{gather}

\begin{figure}[!t]
	\centering
	\includegraphics[width=2.6in]{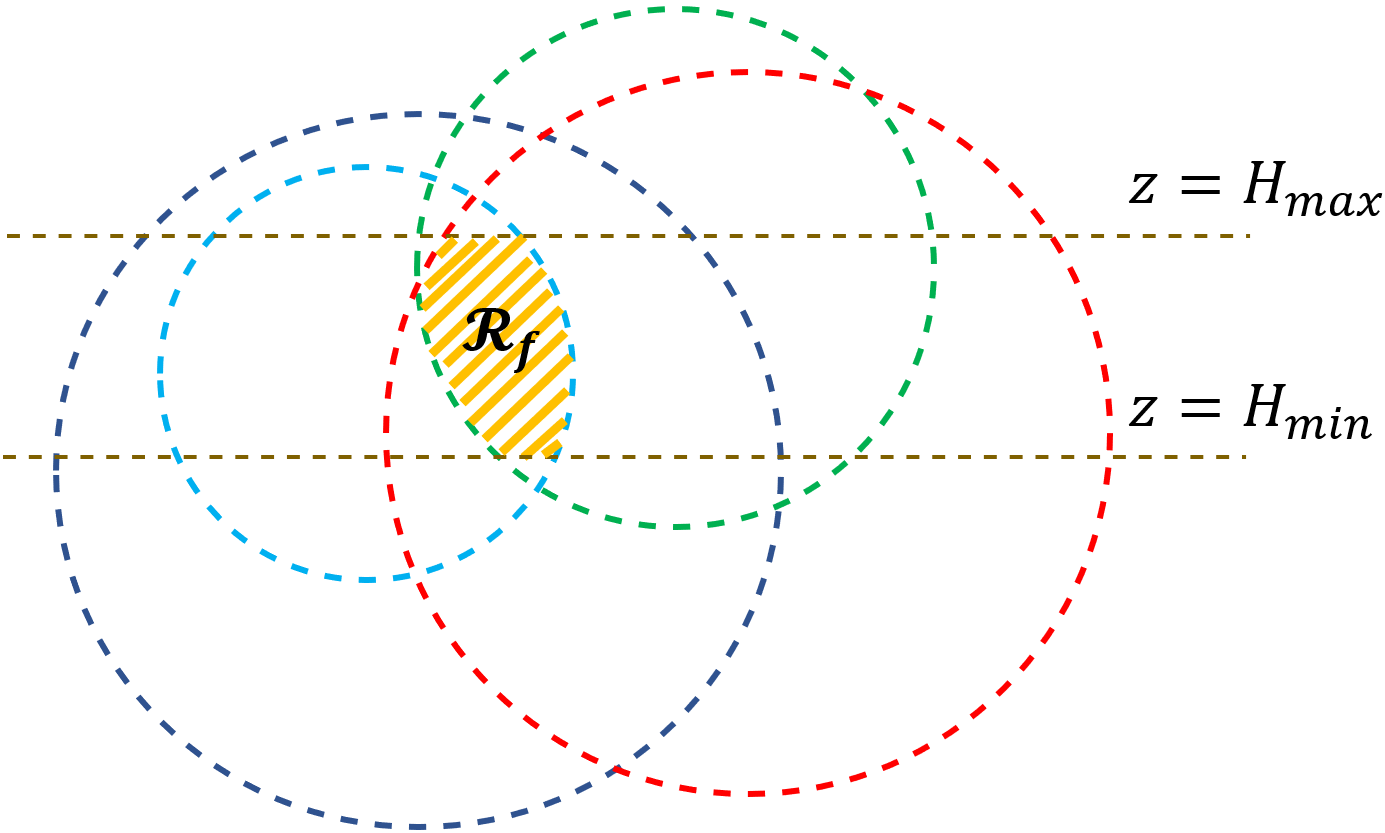}	
	\caption{\textcolor{black}{two-dimensional depiction of feasibility region (hatched in yellow) with respect to the constraints in (\ref{eqn:opt_step4}). }} 
	\label{fig:feasibility_region}
\end{figure}\vspace{0\baselineskip}

\noindent Once (\ref{eqn:5e_equivalent_positioning}) is fulfilled, the constraint (\ref{eqn:problem_formulation_original}e) is automatically fulfilled as well. The left-hand side in (\ref{eqn:5e_equivalent_positioning}) is a constant. Furthermore, the right-hand side in (\ref{eqn:5e_equivalent_positioning}) is strictly decreasing with respect to $d_{F,G}[k]$. Hence, we use bisection method to find the upper bound  $d_{F,G,max}[k]$ such that the inequality
\begin{gather}
d_{F,G}[k]\leq d_{F,G,max}[k],\label{eqn:5e_lesser}
\end{gather}
is equivalent to (\ref{eqn:5e_equivalent_positioning}). The derived upper bound in (\ref{eqn:5e_lesser}) defines the next allowed position of the FlyBS as a sphere centered at the GBS's transmitter and with a radius of $ d_{F,G,max}[k]$.

With the above-provided analysis of the constraints in (\ref{eqn:opt_step3}), we now target the following substitute optimization problem 
\begin{gather}
\operatorname*{max}_{\bm{l_F}[k]}{\;} \sum_{n=1}^{N}C_n[k],\quad\quad\quad\quad\quad \forall k,\label{eqn:opt_step4}\\[-0pt]
\text{s.t.}\quad (\ref{eqn: 5a_in_positioning_subproblem}), (\text{\ref{eqn:problem_formulation_original}b}), (\ref{eqn:4c-4d_combined_1}), (\ref{eqn:5e_lesser}).\nonumber
\end{gather}

\noindent Note that, in later discussions, we refer to the combination of constraints in (\ref{eqn:opt_step4}) as the \textit{feasibility region} at the time step $k$ and we denote it as $\mathcal{R}_f$, i.e.,  $\mathcal{R}_f=\{\bm{l_{F}}| (\ref{eqn: 5a_in_positioning_subproblem}), (\text{\ref{eqn:problem_formulation_original}b}), (\ref{eqn:4c-4d_combined_1}), (\ref{eqn:5e_lesser})\}$). Fig. \ref{fig:feasibility_region} shows a 2D instance of $\mathcal{R}_f$.  \newline

\noindent\textit{C.2) Radial approximation of sum capacity}\vspace{+2mm}


 Now, to tackle the non-convexity of the objective, we propose a radial-basis approximation for the sum capacity. Such approach helps to express the sum capacity as a union of level surfaces determining the direction of  FlyBS's movement towards the optimum position.  In the following, we explain the steps towards the derivation of radial approximation.  Firstly, using (\ref{eqn:friis_tx_rx_Fn}), the log(.) term in (\ref{eqn:1}) is rewritten as 
 
 \small
 \begin{gather}
\log_2{\left(1+\frac{p_{n,F}^R[k]}{\sigma^2_n+p_{n,G}^R[k]}\right)}=\log_2{\left(1+\frac{{Q_{n,F}p_{F,n}^Td_{n,F}^{-\alpha_{n,F}}}}{(\sigma^2_n+ {p_{n,G}^R[k]})}\right)}\label{eqn:radial_step1}
 \end{gather}
 \normalsize
 
  Next, the linear approximation $\log_2(1+X)\approx \frac{X}{\text{ln}(2)}$ is applied to the right-hand side in (\ref{eqn:radial_step1}) to derive a linear expression with respect to  $d_{n,F}^{-\alpha_{n,F}}$. Then, we further derive a linear approximation of $d_{n,F}^{-\alpha_{n,F}}$ with respect to ${d^2_{n,F}}$. In particular, we use the Taylor approximation $(a+X)^k\approx (a+\delta a\xi)^k+k(a+\delta a\xi)^{k-1}(X-\delta a\xi)$ where $\delta=\lfloor \frac{X}{a\xi}\rfloor$, and the parameter $\xi$ determines the accuracy in approximation (smaller $\xi$ leads to smaller error). Using the mentioned approximation for $n\in[1,N]$, we get a sum of quadratic terms in the form of ${d^2_{n,F}}=(X-x_n)^2+(Y-y_n)^2+(Z-z_n)^2$. Since a sum of quadratic terms is also a quadratic expression, the sum capacity $\sum_{n=1}^{N}C_n$ is rewritten as
\vspace{0\baselineskip}
 

\begin{gather}
\sum_{n=1}^{N}C_n[k]\approx W(\bm{p^T},k)-\zeta(\bm{p^T},k){\big|\big|\bm{l_F}[k]-\bm{l_{F,o}}[k]\big|\big|}^2,
\label{eqn:15}
\end{gather}
\normalsize
\noindent where the substitutions $W(\bm{p^T},k)$ and $\zeta(\bm{p^T},k)$ are constants with respect to $\bm{l_F}[k]$. The details about the derivation of (\ref{eqn:15}) are not entirely shown to avoid distraction from the main discussion in this section. Nevertheless, interested readers can refer to \cite{Nikooroo_Pimrc_2022} where more steps of a similar derivation is presented (Appendix A in \cite{Nikooroo_Pimrc_2022} in particular).		



In order to make the approximation in (\ref{eqn:15}) efficient, we derive the approximation  at a position "close" to the actual optimal position, since the objective (sum capacity) is a continuous function. Let $\bm{l_{F,c}}$ denote such a position. We propose to choose $\bm{l_{F,c}}$ by solving an optimization problem derived from (\ref{eqn:problem_formulation_original}) as explained in the following Remark 1. 

\noindent\textbf{Remark 1}: By  using the inequality $ \log_2(1+\frac{1}{ax})\geq \frac{1}{\ln(2)} (\frac{-x}{ax_0^2}+\frac{1}{ax_0}+\ln(1+\frac{1}{ax_0})) $ for arbitrary $a$ and $x$ at any point $x_0$, a lower bound $C_{\text{tot}}^{\text{lb}}$ for the sum capacity is obtained as 
\small
\begin{gather}
\sum_{n=1}^{N}C_n[k]=\nonumber \sum_{n=1}^{N}B_{g_n}\log_2{\left(1+\frac{Q_{n,F}p_{F,n}^T}{d_{n,F}^{\alpha_{n,F}}(\sigma^2_n+ p_{n,G}^R[k])}\right)}\geq
C_{\text{tot}}^{\text{lb}}=\nonumber
\end{gather}
\\
\begin{gather}
 \sum_{n=1}^{N} \frac{B_{g_n}}{\ln(2)}\big(\frac{-Q_{n,F}{d_{n,F}}^{\alpha_{n,F}}}{{(\sigma^2_n p_{n,G}^R[k])d_{n,F}^{2\alpha_{n,F}}[k-1]}}+\frac{Q_{n,F}}{{{(\sigma^2_n+ p_{n,G}^R[k])d_{n,F}^{\alpha_{n,F}}[k-1]}}}\big)\label{eqn:convex_appx}
\end{gather}
\normalsize

\noindent  The right-hand side in (\ref{eqn:convex_appx}) is a concave function with respect to $\bm{l_F}[k]$. Hence,  

\begin{gather}
\bm{l_{F,c}}=\operatorname*{argmax}_{\bm{l_F}[k]}{\;} C_{\text{tot}}^{\text{lb}},\quad \forall k,\label{eqn:opt_step3_part1}\\[-0pt]
\text{s.t.}\quad (\ref{eqn: 5a_in_positioning_subproblem}), (\text{\ref{eqn:problem_formulation_original}b}), (\ref{eqn:4c-4d_combined_1}), (\ref{eqn:5e_lesser}).\nonumber
\end{gather}

\noindent The convex problem in (\ref{eqn:opt_step3_part1}) is solved using CVX. Next, we set $ \bm{l_{F,c}}[k] $ as the reference point for the approximation in (\ref{eqn:15}) as it is a "close" point to the optimal position.  \newline

\noindent\textit{C.3) Solution to FlyBS positioning}\vspace{+2mm}

Now, we elaborate the solution to the FlyBS's positioning. According to (\ref{eqn:15}), the sum capacity increases with a decrease in the distance between $\bm{l_{F,o}}$ and $\bm{l_{F}}$. Thus, the maximum value of sum capacity is achieved at the closest point to $\bm{l_{F,o}}$ that fulfills all the constraints in (\ref{eqn:opt_step4}). According to the discussion in this subsection, each of the constraints (\ref{eqn: 5a_in_positioning_subproblem}), (\ref{eqn:4c-4d_combined_1}), and (\ref{eqn:5e_lesser}) in (\ref{eqn:opt_step4})  limits the FlyBS's position to the border and interior of a sphere and, hence, are convex. Combined with (\text{\ref{eqn:problem_formulation_original}b}), the feasibility region $ \mathcal{R}_f $ for the FlyBS's position is convex. 

Then, the problem of FlyBS's positioning is transformed to 
\begin{gather}
\operatorname*{min}_{\Lambda\in\mathcal{R}_f}{\;} ||\Lambda-\bm{l_{F,o}}[k]||^2,\quad \forall k.\label{eqn:CTO}
\end{gather}

The objective and the domain in (\ref{eqn:CTO}) is convex and, hence, it is solved using CVX.

Once the FlyBS's position $\bm{l_{F}} $ is updated to the solution derived from (\ref{eqn:CTO}), the power allocation $ \bm{p^T} $ is again optimized at the updated position of the FlyBS. Consequently, the updated $ \bm{p^T} $ changes the spheres corresponding to (\ref{eqn: 5a_in_positioning_subproblem}), (\ref{eqn:4c-4d_combined_1}), (\ref{eqn:5e_lesser}) in (\ref{eqn:opt_step4}). Thus, an updated solution to (\ref{eqn:CTO}) is derived. This  optimization of  $ \bm{p^T} $ and $ \bm{l_F} $ is repeated until the FlyBS’s movement at some iteration falls below a given threshold $ \epsilon $ or until the maximum number of iterations is reached.   

\section{Simulations and results}\label{sec:5}

This section provides the details for our adopted simulation scenario followed by  the results and discussions to show superiority of the proposed solution over state-of-the-art.

\subsection{Simulation scenario and models}\label{ssec:sim_scenario}

We assume a 500 m $\times$ 500 m square area with 100 to 600 users initially distributed randomly. The GBS is located  at a distance of 1500 m from the center of the area. We adopt the user's mobility model from \cite{Becvar2022TCOM} where  a half of the users move at a speed of 1 m/s according to random-walk model  and, the other half are randomly divided into six clusters of crowds. A simulation duration of 1200 seconds is assumed. 


A total bandwidth of 100 MHz is divided equally among the users at the access link. The background interference and the noise's spectral density are set to --90 dBm and --174 dBm/Hz, respectively.  Pathloss exponents of $ \alpha_{n,F}=2.3 $, $ \alpha_{n,G}=2.8 $, and $ \alpha_{F,G}=2.1 $ for FlyBS-user, GBS-user, and GBS-FlyBS channels are assumed, respectively \cite{Nikooroo_Globecom_2022}. An altitude range of [100, 300] m and a maximum transmission power limit of $ P_{F,max}^{T}=30 $ dBm is considered for the FlyBS. Also, an altitude of 30 m and a maximum transmission power of $ P_{G,max}^{T}=36 $ dBm (5 W) is assumed for GBS. The results are averaged out over 100 simulation drops.

We benchmark our proposed solution to backhaul-aware sum capacity maximization against the following state-of-the-art schemes: $i$)  \underline{m}aximization of \underline{s}um \underline{c}apacity, referred to as  \textit{MSC}, via FlyBS’s positioning and transmission power allocation at the access link, published in \cite{Nikooroo_Pimrc_2022}, $ii$) \underline{m}inimum \underline{c}apacity \underline{m}aximization, referred to as  \textit{mCM}, via optimization of FlyBS's positioning and transmission power allocation to the users at the access link, published in \cite{Valiulahi2020}, $iii$) \underline{m}aximization of \underline{e}nergy \underline{e}fficiency, referred to as  \textit{EEM}, via transmission power allocation at the access link, as introduced in \cite{Muntaha2021}. Note that the original solution in  \cite{Muntaha2021} does not provide a positioning of the FlyBS, thus, the benchmark scheme EEM is an enhanced version of the solution \cite{Muntaha2021} and the FlyBS's positioning is solved using K-means.

\begin{figure}[!t]
	\centering
	\includegraphics[width=2.2in]{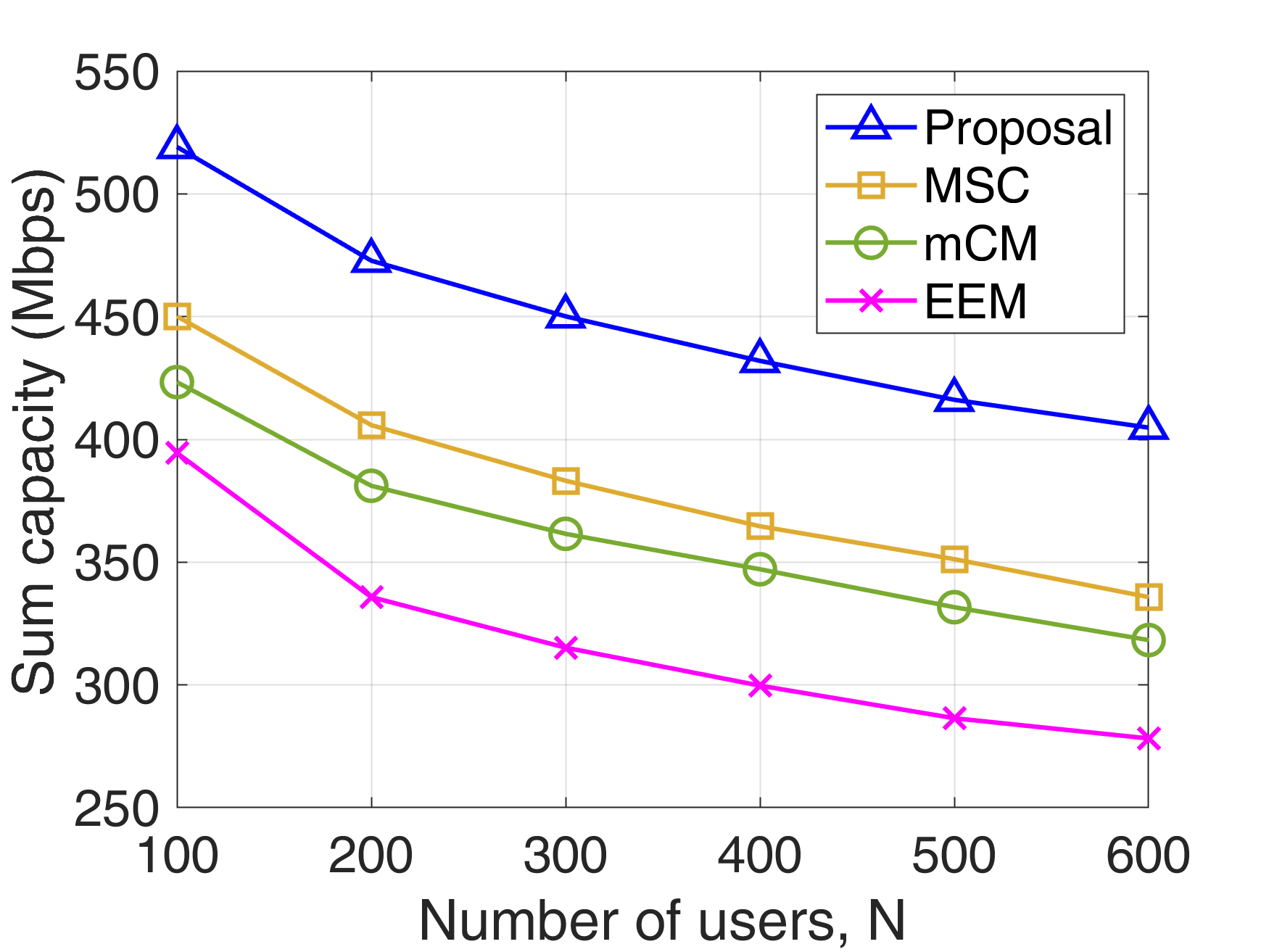}	
	\caption{Sum capacity vs. number of users for $C_{min}=$1 Mbps.} 
	\label{fig:sumC_N}
\end{figure}\vspace{0\baselineskip}

\subsection{Simulation results}\label{ssec:sim_result}
 
In this subsection, we present the simulation results and we discuss the performance of different schemes. 
 
Fig. \ref{fig:sumC_N} shows the sum capacity versus number of users ($ N $) for different schemes. A minimum required capacity of $ C_{min}=1  $ Mbps is assumed for all users.  According to  Fig. \ref{fig:sumC_N}, the sum capacity decreases if more users are served by the FlyBS. This is due to two main reasons 1) the bandwidth allocated to each user becomes smaller, and 2) the FlyBS's total transmission power is divided among more users. Nevertheless, the proposed solution outperforms other schemes in the achieved sum capacity. More specifically, the sum capacity is increased by up to 21\%, 28\%, and 47\% with respect to MSC, mCM, and EEM, respectively.

 \begin{table}
 	\begin{minipage}{0.42\linewidth}		
 		
 		\centering
 		\includegraphics[width=1.8in]{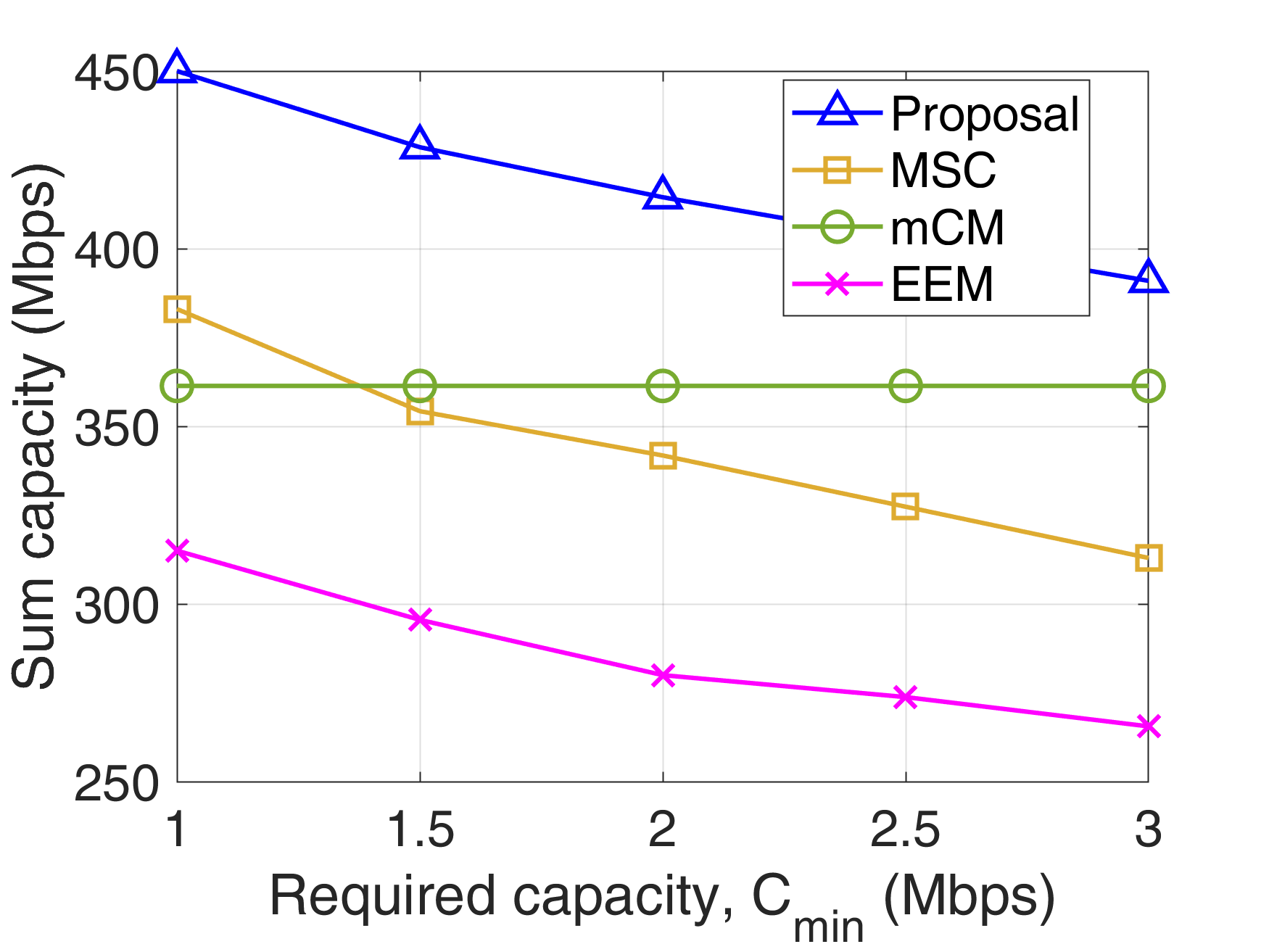}
 		\captionof{figure}{\textcolor{black}{Sum capacity vs. $C_{n,min}$ for $N=$~300.}}
 		\label{fig:sumC_Cmin_N=300}
 	\end{minipage}
 	\hspace{0.6cm}
 	\begin{minipage}{0.42\linewidth}
 		\centering
 		\includegraphics[width=1.8in]{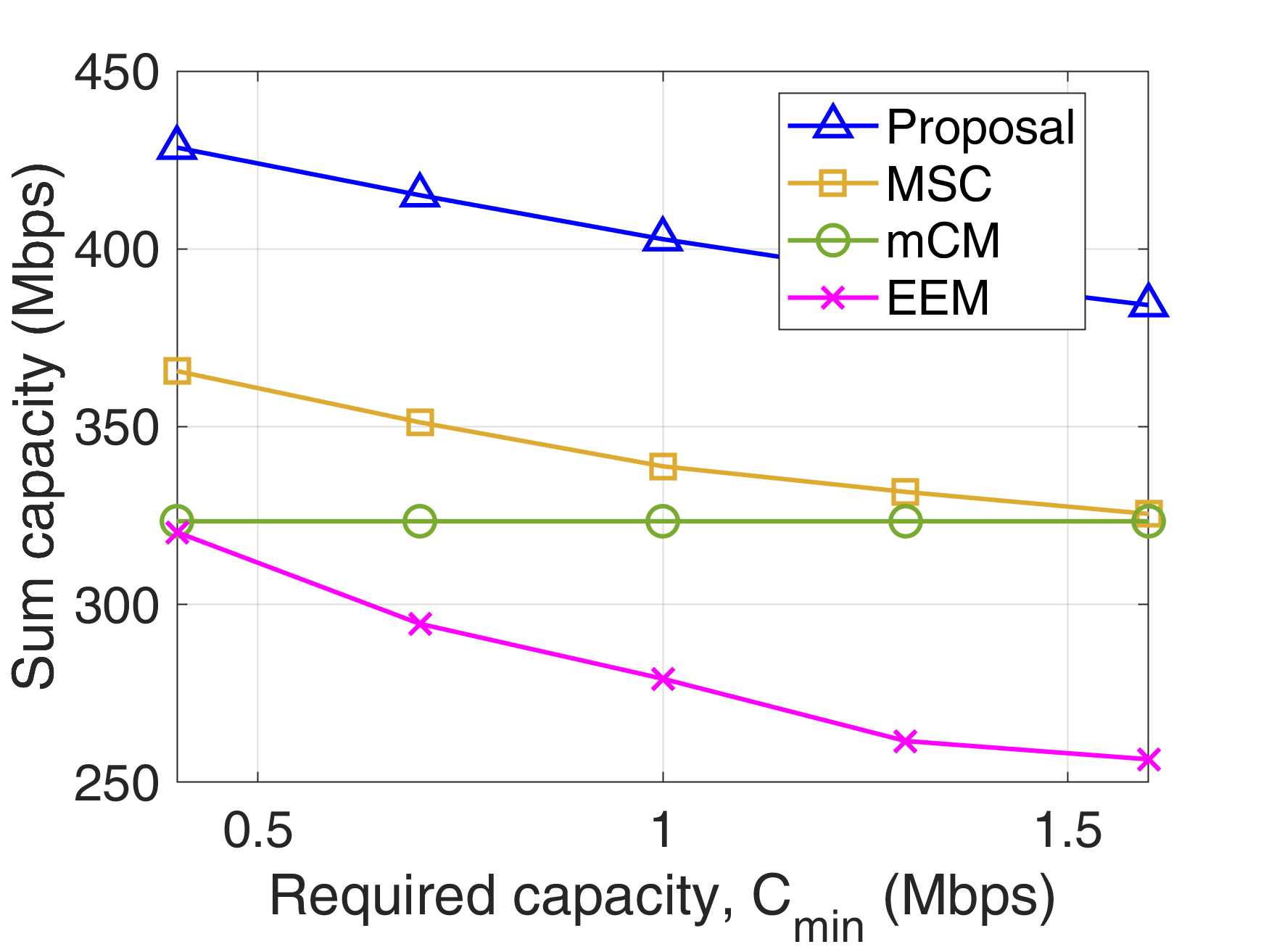}	
 		\captionof{figure}{\textcolor{black}{Sum capacity vs. $C_{n,min}$  for $N=$~600.}} 
 		\label{fig:sumC_Cmin_N=600}	
 	\end{minipage}
 \end{table}

Next, Figs. \ref{fig:sumC_Cmin_N=300} and \ref{fig:sumC_Cmin_N=600} show the impact of minimum user's capacity $C_{min}$ on the sum capacity for $ N=$~300 and $ N=$~600, respectively. 
The maximum depicted $ C_{min} $ represents the largest $ C_{min} $ for which a feasible solution exists. However, the value of $ C_{min} $ in \textit{mCM} is not set manually and beforehand, as it is directly derived by the scheme itself (which maximizes the minimum capacity). 
Hence, the sum capacity is constant in Figs. \ref{fig:sumC_Cmin_N=300} and \ref{fig:sumC_Cmin_N=600}. However, for the proposed solution, MSC, and EEM, increasing $ C_{min} $ reduces the sum capacity. This is because increasing $ C_{min} $ leads to a tighter feasibility region according to (\ref{eqn: 5a_in_positioning_subproblem}) and, hence, it  limits the FlyBS's movement to maximize the sum capacity. 
 The proposed solution increases the sum capacity with respect to \textit{MSC}, \textit{mCM}, and \textit{EEM} by 24\%, 25\%, and 49\%, respectively, for $ N= $~300, and by 19\%, 33\%, and 49\%, respectively, for $ N= $~600.

Next, we also demonstrate the fast convergence of our proposed iterative algorithm in Figs. \ref{fig:sumC_iter_N=300} and \ref{fig:sumC_iter_N=600} by showing an evolution of the sum capacity over iterations the alternating optimization of transmission power allocation and FlyBS's positioning. Note that, the benchmark schemes mCM and EEM are not iterative and, hence, their sum capacity is constant and they are shown in the Figs. \ref{fig:sumC_iter_N=300} and \ref{fig:sumC_iter_N=600} only to show their performance. The proposed solution converges very fast and in only few iterations. This confirms that the iterative manner of the proposed solution does not limit its feasibility and practical application. Note that, although the mCM scheme outperforms our proposal in the first iteration in Fig. \ref{fig:sumC_iter_N=600}, only the converged results should be subject to comparison as the performance at early iterations can be greatly impacted by the initialization of FlyBS's position and power allocation.


\begin{table}
	\begin{minipage}{0.44\linewidth}		
		
		\centering
		\includegraphics[width=1.8in]{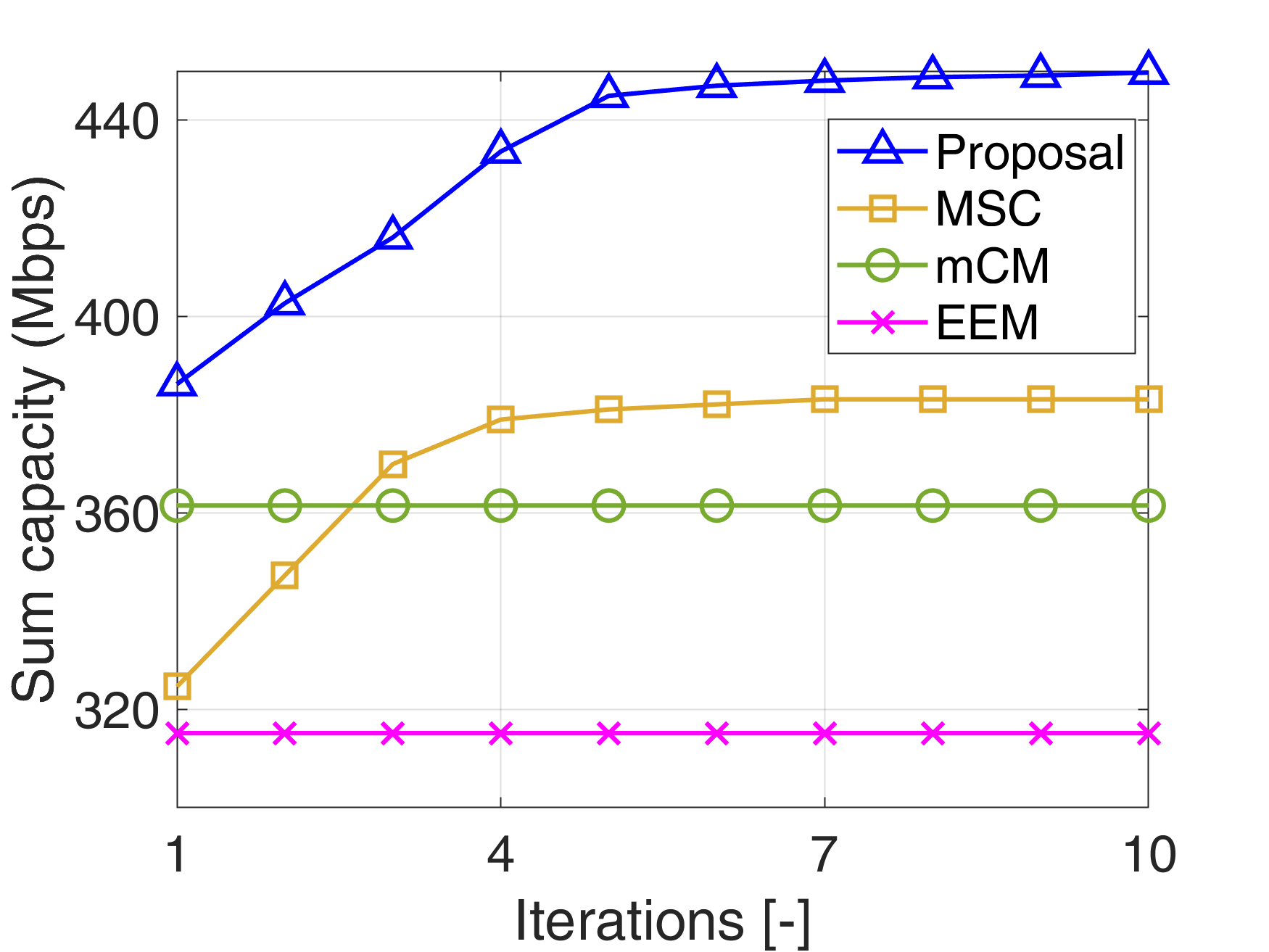}
		\captionof{figure}{Convergence of the proposed scheme for $N=$300.}
		\label{fig:sumC_iter_N=300}
	\end{minipage}
	\hspace{0.5cm}
	\begin{minipage}{0.44\linewidth}
		\centering
		\includegraphics[width=1.8in]{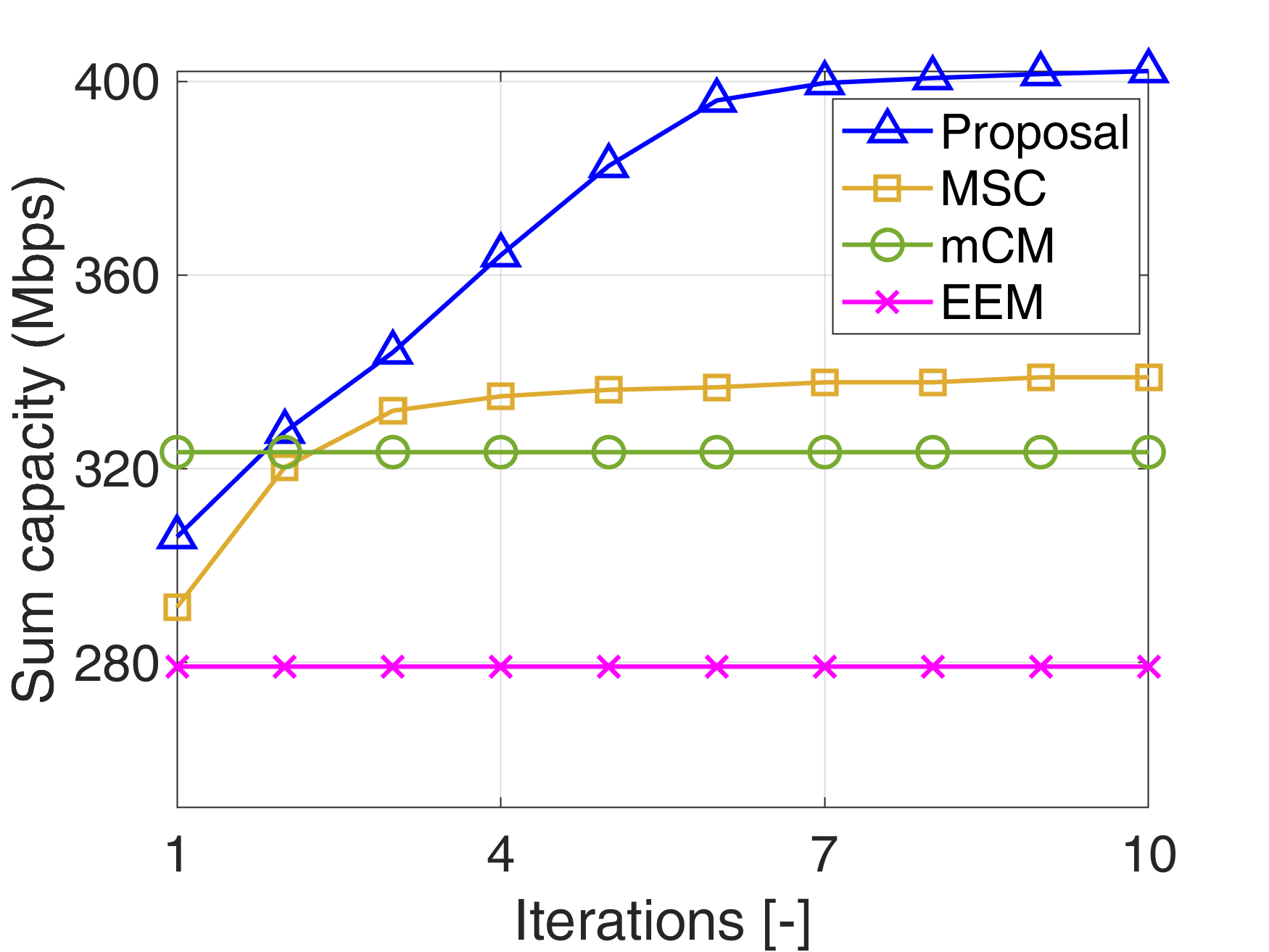}	
		\captionof{figure}{Convergence of the proposed scheme for $N=$600.} 
		\label{fig:sumC_iter_N=600}	
	\end{minipage}
\end{table}

\section{Conclusions}

In this paper, we have provided an analytical approach to maximize the sum capacity via a positioning of the FlyBS, allocation of transmission power to the backhaul channels,  and an allocation of the transmission power to the users at the access channel. The problem is constrained by the minimum required instantaneous capacity to each user and practical real world limitations of the FlyBSs. We have shown that the proposed solution enhances the sum capacity by tens of percent compared to state-of-the-art works. In the future work, a scenario with multiple FlyBSs should be studied \textcolor{black}{along with related aspects, such as a management of interference among FlyBSs and an association of  users to  FlyBSs.}





\end{document}